\documentstyle[preprint,revtex]{aps}
\begin{document}
\draft
\begin{title}
Theory of Interacting Parallel Quantum Wires
\end{title}
\author{Yinlong Sun and George Kirczenow}
\begin{instit}
Department of Physics, Simon Fraser University, Burnaby, BC, Canada
V5A 1S6\\
Phone: (604) 291-2502 or (604) 574-3848, Fax: (604) 291-3592\\
E-mail: sun@sfu.ca or kirczeno@sfu.ca
\end{instit}
\begin{flushleft}
Index number: 7335 Mesoscopic systems
\end{flushleft}

\newpage

\begin{abstract}
\\
ABSTRACT

We present self-consistent numerical calculations of the electronic
structure of parallel Coulomb-confined quantum wires, based on the
Hohenberg-Kohn-Sham density functional theory of inhomogeneous
electron systems.
We find that the corresponding transverse energy levels of two
parallel wires lock together when the wires' widths are similar and
their separation is not too small.
This energy level locking is an effect of Coulomb interactions and of
the the density of states singularities that are characteristic of quasi-
one-dimensional Fermionic systems.
In dissimilar parallel wires level lockings are much less likely to
occur.
Energy level locking in similar wires persists to quite large wire
separations, but is gradually suppressed by inter-wire tunneling
when the separation becomes small.
Experimental implications of these theoretical results are discussed.
\end{abstract}
\newpage
\section{Introduction}
\label{sec:1}
Nanostructures fabricated from semiconductor heterostructures have
stimulated many studies in recent years.\ \cite{Ulloa}
Various techniques \cite{Ulloa,Roukes} such as electron-beam
lithography, ion-beam exposure, and etching have been developed to
confine the two-dimensional
electron gas (2DEG) in heterostructures so as to form quantum wires
and constrictions with characteristic dimensions on the 100~nm scale.
There have been many theoretical studies of the electronic structure
of single quantum wires \cite{Lai,Laux-86,Laux-
88,Davies,Glazman,Nakamura,Sun-93}.
For multiple parallel wires, theoretical studies of transport in models
of non-interacting electrons \cite{Avishai,Castano} and interacting
electrons \cite{Vasilopoulos} have been published, and electronic
structure calculations of coupled quantum wires have also begun to
appear \cite{Ravaioli,Sun-94}.
Recently, it has been demonstrated \cite{Sun-94}, using the density
functional theory of Hohenberg Kohn and Sham, that the transverse
levels of parallel quantum wires can lock together under certain
conditions.
As a novel phenomenon, the effect of energy level locking deserves
further study and should have interesting implications for
experiments.

The purpose of this paper is to present a systematic theoretical study
of the electronic structures of parallel quantum wires, with emphasis
on the effect of energy level locking.
The quantum wires we consider are of the Coulomb-confined type
\cite{Sun-93}, and the calculations are performed self-consistently.
In Sec.\ \ref{sec:2}, we review briefly the formalism of the
Hohenberg-Kohn-Sham
density functional theory, and its application to the present quantum
wire
system. In Sec.\ \ref{sec:3}, we present the results of calculations for
two
parallel quantum wires in different situations.  In similar wires, the
effect
of energy level locking is found to be favored and the origin of the
effect is
charge transfer between the wires.  In dissimilar wires, however, the
electronic
structure is quite different, and energy level locking is much less
likely to
occur.  Energy level locking in similar wires remains strong for quite
large
wire separations, because the energy cost of the charge transfer
depends
logarithmically on the wire separation. But the effect of energy level
locking
is gradually suppressed by the tunneling between wires when the
wire separation
becomes small. In Sec.\ \ref{sec:4}, we present a brief discussion of
the
experimental implications of these theoretical results.

The structure of the parallel quantum wires that we consider is
shown in
Fig.\ \ref{model}. Electrons are confined to the x-y plane, which may
represent a semiconductor heterointerface. Two uniform positive
ribbons A and B, representing regions of donors in a semiconductor
heterostructure, extend infinitely in the y-direction, offset from
the x-y plane by a distance $d$.  The ribbons have widths of $w_{a}$
and $w_{b}$, respectively, and a separation $s$.  The whole system is
charge-neutral, and is embedded in a uniform dielectric.  The 2DEG
has a uniform Fermi energy and is confined laterally by the
self-consistent effective potential that includes the Coulomb
interaction and the many-body effects of exchange and correlation.

\section{Density Functional Theory}
\label{sec:2}
The Hohenberg-Kohn-Sham density functional theory
\cite{Hohenberg,Kohn} provides an accurate treatment of the effects
of exchange and correlation for the ground state properties of
inhomogeneous electron systems.
In this theory, the effective Schr\"{o}dinger equation for an electron
of a 2DEG is
\begin{equation}
- {\hbar^2 \over 2m^{*}} \left[ {d^2 \over {dx^2}} + {d^2 \over {dy^2}}
\right] \Psi_{\ell k}(x,y) + V_{eff} \left[ n;x,y \right] \Psi_{\ell k}(x,y)
= E_{\ell k} \Psi_{\ell k}(x,y),
\label{Schodinger2}
\end{equation}
where $\Psi_{\ell k}(x,y)$ and $E_{\ell k}$ are the eigenfunction and
eigen energy, respectively. $V_{eff}[n;x,y]$ is the effective potential
energy, which is a functional of the two-dimensional electron density
$n(x,y)$. In the local density approximation,
\begin{equation}
V_{eff} \left[ n;x,y \right] = V_{C}(x,y) + E'_{xc}[n] = V_{C}(x,y) + {d
\over dn} (n \varepsilon_{xc}[n] ), \label{Veff}
\end{equation}
where $V_{C}$ is the Coulomb energy and $\varepsilon_{xc} =
\varepsilon_{x} + \varepsilon_{c}$, with $\varepsilon_{x}$ and
$\varepsilon_{c}$ being the exchange and correlation energies per
electron, respectively.
For a 2DEG, the exchange energy $\varepsilon_{x}$ has the following
analytic form \cite{Stern}
\begin{equation}
\varepsilon_{x}[n] = -{8 \sqrt{2} \over 3 \pi r_{s} }= -{8 a_{0}^{*}
\over 3 \pi} \sqrt{2 \pi n},
\label{exchange}
\end{equation}
where $a_{0}^{*}$ is the effective Bohr radius.
The correlation energy $\varepsilon_{c}$ has been calculated
numerically \cite{Jonson,Tanatar}.

For the two parallel quantum wires shown in Fig.\ \ref{model}, the
two-dimensional density of donors in the charged ribbons is
described by
\begin{equation} n_{d}(x)= \left \{  \begin{array}{ll}
\sigma_{a} & \mbox{if $-s/2-w_{a}<x<-s/2$} \\
\sigma_{b} & \mbox{if $s/2<x<s/2+w_{b}$} \\
0 & \mbox{otherwise}
\end{array}
\right.
\end{equation}
Because the system is uniform in the y-direction,
equation~\ref{Schodinger2} reduces to the one-dimensional form
\begin{equation}
- {\hbar^{2} \over 2m^{*}} {d^{2} \over dx^{2}} \Phi_{\ell}(x) + V_{eff}
\left[ n;x \right] \Phi_{\ell}(x) = E_{\ell} \Phi_{\ell}(x),
\label{Schodinger1}
\end{equation}
where
\begin{equation}
\Psi_{\ell k}(x,y) =\Phi_{\ell}(x) e^{iky},  E_{\ell k} = E_{\ell} + {
\hbar^{2} k^{2} \over 2m^{*}}.
\label{separation}
\end{equation}
In equations~\ref{Schodinger1} and \ref{separation}, $\Phi_{l}(x)$
and $E_{l}$ are the transverse eigenfunction and eigen energy,
respectively, and $k$ is the wave number in the y-direction.
At zero temperature, the two-dimensional electron density $n(x)$
relates to the transverse wave functions through
\begin{equation}
n(x)=2 \sum_{\ell k} | \Psi_{\ell k}(x,y)|^{2} = {2 \over \pi} \sqrt
{2m^{*} \over \hbar^{2}} \sum_{\ell, E_{\ell} \leq  E_{F}} \sqrt{E_{F}-
E_{\ell}}  |\Phi_{\ell}(x)|^{2}.
\label{density}
\end{equation}
After performing the integral over $x$ on both sides of equation
\ref{density}, we obtain
\begin{equation}
N = \int_{-\infty}^{\infty} dx n(x) = {2 \over \pi} \sqrt {2m^{*} \over
\hbar^{2}}
\sum_{\ell, E_{\ell} \leq  E_{F}} \sqrt{E_{F}-E_{\ell}},
\label{Fermi}
\end{equation}
where $N$ is the linear electron density in the y-direction.
Finally, choosing the potential at $x = \pm \infty$ to be zero, the
Coulomb energy in equation \ref{Veff} is given by
\begin{equation}
V_{C}(x)={e^{2} \over 4 \pi \epsilon_{0} \epsilon } \int_{-
\infty}^{\infty} dx' \int_{-\infty}^{\infty} dy' \left \{
{n(x') \over \sqrt{(x-x')^{2}+y'^{2}} }-{n_{d}(x') \over \sqrt{(x-
x')^{2}+y'^{2} + d^{2}} } \right \}.
\label{VC}
\end{equation}

The inputs for the numerical calculations are the geometric
parameters $w_{a}$, $w_{b}$, $s$, and $d$, and the donor densities
$\sigma_{a}$ and $\sigma_{b}$.
Using equations \ref{Veff}, \ref{Schodinger1}, \ref{density}, and
\ref{Fermi}, and \ref{VC}, the transverse wavefunction
$\Phi_{\ell}(x)$, transverse level $E_{\ell}$, Fermi energy  $E_{F}$,
and electron density $n(x)$ were calculated self-consistently.
In our calculations, we used the effective mass $m^{*}$=0.067 and
the dielectric constant $\epsilon$=12.5, corresponding to the values
for GaAs.

\section{Numerical Results and Discussion}
\label{sec:3}
In this section, we present the self-consistent numerical results at
zero temperature for two parallel quantum wires in different
situations.
The calculated electronic structures of similar and dissimilar parallel
wires are presented in separate subsections, because their features
are very different.
Some qualitative analyses are made following the numerical results.
A theoretical summary is given at the end of this section.
\subsection{Similar parallel wires}
The calculated electronic structure of two similar parallel quantum
wires is shown in
Fig.\ \ref{similar}.  The widths of wires are $w_{a}$ = 190~nm and
$w_{b}$ = 200~nm, respectively.  The other geometric parameters
are
$s$ = 200~nm and $d$ = 20~nm. The donor densities in the two
charged
ribbons are kept the same,  $\sigma_{a}$ = $\sigma_{b}$ = $\sigma$.
Since the system is charge-neutral overall, $\sigma$ can also be
regarded as an electron filling parameter. In Fig.\ \ref{similar}, the
solid lines are the six lowest transverse energy levels $E_{\ell}$,
and the dashed line is the Fermi energy $E_{F}$.  The energy levels
are labelled A or B, according to whether they belong primarily to
wire A or B, respectively, as determined by inspection of the
calculated eigenfunctions.

Notice that when the Fermi energy rises up through the lower of a
pair of adjacent energy levels with increasing $\sigma$, the
(algebraic) slope of the lower energy curve of the pair increases
while that of the upper curve decreases.
Thus the two corresponding energy levels are brought closer
together.
Levels 3 and 4 lock together at the Fermi energy, while the gap
between levels~5 and 6 narrows by a factor of about 5.
These effects are referred to as ``energy level locking''.
In Fig.\ \ref{similar}, we observe three features that characterize
energy
level locking. The first is that the effect is associated with the Fermi
energy
crossing the corresponding levels. The second is that the locked
levels tend to
remain together. The third feature is that the sequence of locked
levels
remains unchanged throughout.  (For the lockings in Fig.\
\ref{similar}, the
level belonging to wire B is always lower than the level belonging to
wire A.)

The energy level locking is caused by a charge imbalance that occurs
between the wires and modifies the Coulomb potentials when a
transverse level begins to fill.
The origin of the charge imbalance is the $E^{-1/2}$ density of states
singularity at the bottom of a subband, characteristic of quasi-one-
dimensional systems.
Note that a transverse level $E_{\ell}$ corresponds to the bottom of
subband $E_{\ell k}$, as reflected in equation \ref{separation}.
When the Fermi energy rises up through $E_{\ell}$, because of the
density of states singularity, most of the added electrons go into
subband $E_{\ell k}$.
Thus the wire to which $E_{\ell}$ mainly belongs acquires an excess
of electrons, and a charge imbalance occurs.
Such a charge imbalance, through the Coulomb interaction, shifts the
self-consistent electrostatic potential and the transverse energy
levels of the wire with the excess (deficiency) of electrons upwards
(downwards) significantly, favoring energy level locking.
Energy levels that lock together do not separate immediately when
the Fermi energy rises above them, because their density of states
singularities almost coincide, which inhibits further changes of the
charge differential.

Some features of the electronic structure in Fig.\ \ref{similar} can be
understood qualitatively as a competition between a charge
imbalance and
inter-wire quantum hybridization. The inter-wire quantum
hybridization acts to
separate the adjacent levels, and thus opposes energy level locking.
In
Fig.\ \ref{similar}, the separations between levels 3 and 4 and
between levels
5 and 6 narrow markedly when the Fermi energy crosses them. This
implies that
the charge imbalances are the dominant factor. However, the energy
gap between
levels 5 and 6 remains fairly large. This is because inter-wire
quantum
tunneling and therefore hybridization is more significant for higher
levels.  A
charge imbalance occurring in a high level induces an electrostatic
potential
that acts like an external field on lower levels. Thus this charge
imbalance
can cause the corresponding pairs of lower levels to lock together or
even
anticross. In this way, in Fig.\ \ref{similar}, when level 5 begins to
fill,
level 3 and 4 are brought together significantly. On the other hand,
levels 1
and 2 anticross twice (indicated by the arrows), corresponding to the
Fermi
energy crossing level 5 and 6, respectively.

To show that a charge transfer does occur when a transverse level
begins to fill, we present the following simple argument.
Consider the situation where all populated levels are tightly bound,
so that the wavefunction overlaps between the different wires are
negligible.
Suppose when $\sigma$ = $\sigma_{0}$, the Fermi energy is
$E_{F}(\sigma_{0})$ = $E_{3}$, (see
Fig.\ \ref{parabolic}). According to equation~\ref{Fermi}, the ratio of
the
numbers of electrons in wires A and B is given by \begin{equation}
r(\sigma_{0})={N_{a}(\sigma_{0}) \over N_{b}(\sigma_{0})}
={\sqrt{ E_{F}(\sigma_{0})- E_{2}} \over
 \sqrt{ E_{F}(\sigma_{0})- E_{1}} }.
\label{ratio1}
\end{equation}
For simplicity, let us assume that the potentials remain unchanged
although $\sigma$ increases.
In other words, the increase of $E_{F}$ is completely caused by the
increase of $\sigma$.
Suppose that when $\sigma$ increases from $\sigma_{0}$ by a small
amount $\Delta \sigma$, $E_{F}$ increases by a small amount $\Delta
E$ so that $E_{F}$ locates above level~3 but still below level~4.
This situation is shown in Fig.\ \ref{parabolic}.
Then, at $\sigma$ = $\sigma_{0}$ + $\Delta \sigma$, the electron
linear densities in wires A and B are
\begin{equation}
N_{a}(\sigma_{0}+\Delta \sigma) = N_{a}(\sigma_{0})+ {1 \over \pi}
\sqrt{2m^{*} \over \hbar^{2}} {\Delta E \over
\sqrt{E_{F}(\sigma_{0})- E_{2}}}
\label{Na}
\end{equation}
and
\begin{equation}
N_{b}(\sigma_{0}+\Delta \sigma) = N_{b}(\sigma_{0})+ {1 \over \pi}
\sqrt{2m^{*} \over \hbar^{2}} {\Delta E \over
\sqrt{E_{F}(\sigma_{0})- E_{1}}} + {2 \over \pi}
\sqrt{2m^{*} \over \hbar^{2}} \sqrt{\Delta E },
\label{Nb}
\end{equation}
respectively.
The last term in equation \ref{Nb} is due to level~3.
The other two terms containing $\Delta E$ in equations \ref{Na} and
\ref{Nb} are associated with the 2nd and 1st levels, respectively.
Because $\Delta E$ is very small, the term due to level~3 in equation
\ref{Nb} dominates the other terms.
Keeping to the lowest order in $\Delta E$, $r$ becomes
\begin{equation}
r(\sigma_{0} +\Delta \sigma) = r(\sigma_{0}) - r(\sigma_{0})
{\sqrt{ \Delta E} \over \sqrt{ E_{F}(\sigma_{0})- E_{1}} }.
\label{ratio2}
\end{equation}
This equation implies that a charge imbalance occurs when a new
level begins to fill.

If $\sigma$ increases from $\sigma_{0}$ by $\Delta \sigma$ so that
$E_{F}$ rises above both levels~3 and 4, another term due to level~4
should be added to the expression of $r$
\begin{equation}
r(\sigma_{0} +\Delta \sigma) = r(\sigma_{0}) - r(\sigma_{0})
{\sqrt{ \Delta E} \over \sqrt{ E_{F}(\sigma_{0})- E_{1}} }
+ r(\sigma_{0})
{\sqrt{ \Delta E'} \over \sqrt{ E_{F}(\sigma_{0})- E_{2}} },
\label{ratio3}
\end{equation}
where $\Delta E' = E_{F}(\sigma_{0} +\Delta \sigma) - E_{4}$.
In similar parallel wires, because $E_{1} \sim E_{2}$ and $E_{3} \sim
E_{4}$, the two terms in equation \ref{ratio3} tend to cancel each
other.
Because of this cancellation, when $\sigma$ increases, further
differential charging is inhibited and thus the locked levels tend to
stay together.

Obviously, the above argument also holds for situations when higher
levels are crossed by the Fermi energy.
In the self-consistent calculations, however, the electrostatic
potentials and transverse levels are actually affected by the charge
imbalance and move in response to it.
(These electrostatic level shifts are in fact responsible for the energy
level locking.)
Because of this electrostatic response, the self-consistent charge
imbalance is not as large as that given by equations \ref{ratio2} and
\ref{ratio3}.
In Fig.\ \ref{transfer}, we display the calculated ratio $r$ for the
range of
$\sigma$ in which the Fermi energy crosses the 5th and 6th
transverse levels.
Notice that $r_{0} = w_{a}/w_{b}$ = 0.95 corresponds to perfect
charge balance
between the wires. Our calculation shows that $r$ oscillates about
$r_{0}$; its
drop near $\sigma = {\rm 1.8 \times 10^{10}~cm^{2} }$ and rise
beginning at
$\sigma = {\rm 2.0 \times 10^{10}~cm^{2} }$ are due to the 5th and
6th
transverse levels beginning to fill, respectively.

In Fig.\ \ref{similar}, another feature of the electronic structure is
that all
transverse energy levels first decrease and then increase in energy
while
$\sigma$ increases. This behavior results from the competition
between the
Coulomb energy and the exchange-correlation energy. According to
equation
\ref{exchange}, $\varepsilon_{x} \propto -n^{1/2}$.  The correlation
energy
$\varepsilon_{c}$ also increases negatively with $n$, but slower than
$\varepsilon_{x}$. \cite{Jonson,Tanatar} The Coulomb energy has no
strict
power-law dependence on the electron density, because the electron
density
appears in the integral in equation \ref{VC}. However, when all
populated
transverse levels are tightly bound, electrons distribute mainly
within the
potential wells and thus approximately $V_{C} \propto n$. At low
electron
densities, the exchange-correlation energy dominates the Coulomb
energy. In
fact, the exchange-correlation energy confines electrons so tightly
that the
Coulomb energy is overall positive. \cite{Sun-93} Because the
exchange-correlation energy dominates the Coulomb energy at low
densities,
increasing $\sigma$ results in further lowering of all transverse
levels.
However, when the electron density is increased sufficiently, the
Coulomb
energy becomes more important, thus causing the energy curves to
become flat
and then to rise gradually.

The competition between the Coulomb energy and the exchange-
correlation energy also affects the energy level locking in an
important way.
According to equations \ref{Veff} and \ref{exchange}, when there is
a small increase of the electron density $\Delta n$, the variation of
contribution from the exchange energy
itself to the total effective potential energy is
\begin{equation}
\Delta E'_{x} = -{4 a_{0}^{*} \Delta n \over \sqrt{2 \pi n}}.
\end{equation}
However, $\Delta V_{C} \propto \Delta n$, that is, approximately
independent of $n$.
When $n$ is extremely low, $\Delta E'_{x}$ may dominate $\Delta
V_{C}$.
Then, since $\Delta E'_{x}$ differs in sign from $\Delta V_{C}$, a
charge
imbalance will not result in energy level locking at extremely low
densities.
Therefore, energy level locking in similar wires also requires a
sufficient
electron density so that the Coulomb energy is dominant.

To study the role of inter-wire separation on energy level locking, we
have calculated the electronic structure as a function of the wire
separation $s$.
The result is shown in Fig.\ \ref{wireseparation}.
The parameters used were $w_{a}$ = 190~nm, $w_{b}$ = 200~nm,
$d$ = 20~nm, and $\sigma_{a} = \sigma_{b} = {\rm 2.0 \times
10^{10}~cm^{2} }$, which are the locking of levels~5 and 6 in
Fig.\ \ref{similar}. The solid lines are the eight lowest levels and the
dotted
line is the Fermi energy. On the right side of this figure, the solid
lines
ending with solid circles and the dashed lines ending with open
circles
correspond to the energy levels of isolated wires A and B,
respectively.

Compared to the gaps between corresponding levels of the isolated
wires, we observe strong effects of energy level locking for large
wire separation.
When $s$ is small, however, the inter-wire tunneling becomes
strong, which causes the energy levels to become well separated
because of hybridization.
To illustrate the role of the tunnelling when $s$ becomes small, we
show in
Fig.\ \ref{wavefunction} the wave functions of the lowest four levels
for a
wire separation $s$ = 100~nm.
Notice that the widths of the electrostatically confined quantum
wires are somewhat larger than the widths of the ribbons of positive
charge that confine the electrons.
In case shown there is significant tunneling between the wires and
the higher energy
 wavefunctions have similar amplitudes in both wires. In Fig.\
\ref{wireseparation}, the lower
pairs of levels show more tendency to lock together, because the
electrons of low levels experience a higher barrier between wires.
Finally, we should point out that levels~3 and 4 are closer than
levels~1 and 2 for large $s$, because, at this particular value of
$\sigma$, the order of levels~1 and 2 is reversed (check the
wavefunctions or refer to Fig.\ \ref{similar}).

An interesting feature of Fig.\ \ref{wireseparation}  is that even for
the large wire separation $s$ of 800 nm, the gaps between the paired
levels of the two interacting wires are still much smaller than the
gaps between the corresponding levels of isolated wires. Thus the
energy level locking found in the present model is a quite long-range
effect. The reason for this is that the Coulomb energy cost of the
charge transfer between infinite parallel wires depends
logarithmically on the distance between the wires (for large $s$),
and is thus insensitive to the wire spacing. On the other hand, the
tunneling between wires that opposes the energy level locking
decreases exponentially as $s$ increases.

\subsection{Dissimilar parallel wires}

The electronic structures of dissimilar parallel quantum wires are
quite different from those of similar parallel quantum wires.
In Fig.\ \ref{dissimilar1}, we present the calculated electronic
structures
for  $w_{a}$ = 100~nm and $w_{b}$ = 200~nm. The other geometric
parameters are
$s$ = 200~nm and $d$ = 20~nm, and $\sigma_{a}$ = $\sigma_{b}$ =
$\sigma$. The
solid lines are the few lowest transverse energy levels, and the
dotted line
is the Fermi energy. The levels are again labelled A or B according to
which
wire they primarily belong to.

Generally speaking, in dissimilar parallel wires, the transverse levels
of the two wires are well separated from each other.
When the Fermi energy crosses a transverse level, an abrupt charge
imbalance occurs for the same reason as in similar wires.
The charge imbalance can significantly twist the curves of transverse
levels, but is not sufficient to lock two levels together.
The level twists are seen where the Fermi energy crosses the 3rd
level in
Fig.\ \ref{dissimilar1}. However, if a pair of levels happen to be
close when the Fermi energy crosses them, they can be squeezed
together significantly by the charge imbalance. This is reflected in
the crossing between levels 4 and 5 (indicated by an arrow). This
energy crossing, however, is a case of ``anticrossing'' instead of
level locking, because the level sequence reverses. The anticrossing
is associated with the asymmetry of the two wires. In this case, the
charge imbalance narrows the energy gap at the anticrossing, but the
levels then separate quickly.

Another way to study two dissimilar parallel wires is by varying the
donor density of one wire, while fixing the donor density of the other
wire.
Such a case is shown in Fig.\ \ref{dissimilar2}.
Here, $\sigma_{b}$ is varied while $\sigma_{a}$ is fixed at ${\rm 1.5
\times 10^{10}~cm^{2} }$.
The parameters used here are $w_{a}$ = 190~nm, $w_{b}$ = 200~nm,
$s$ = 200~nm, and $d$ = 20~nm.
When $\sigma_{b}$ increases, the Fermi energy also increases for the
most part.
To keep the Fermi energy the same in both wires, some electrons
must transfer from wire B to wire A.
These excess electrons cause wire A to have a net negative charge,
and thus its energy levels rise with increasing $\sigma_{b}$.
On the other hand, because wire B is deficient of electrons, its levels
fall.
Since the levels in wire A increase with the Fermi energy, their
trajectories are similar to that of the Fermi energy.
The Fermi energy is therefore unlikely to cross the levels of wire A.
In other words, the Fermi energy can cross only one level at a time.

\subsection{Summary}

Based on above discussion, we summarize the conditions for energy
level locking as follows.

1) The system should be quasi-one-dimensional and consist of
parallel subsystems with Coulomb interaction.

2) The subsystems should be similar.

3) The electron density should not be too small.

4) The separation between the subsystems should be large enough
for tunnelling between them to be weak.

It should be noted that the quasi-one-dimension condition is
essential for energy level locking.
To see this, let us consider a quasi-two-dimensional system, that is,
an electron system that is confined in two quantum wells.
Because there is no density of states singularity in the two-
dimensional system, analogous to equation \ref{Nb}, we now have
\begin{equation}
\Delta N_{b}= {m^{*} \over \pi \hbar^{2}} \sum_{\ell, E_{\ell} < E_{F}}
\Delta E,
\label{Nb2}
\end{equation}
where $\Delta N_{b}$ is the variation of the area electron density in
well B.
This implies that all populated levels contribute to $\Delta N_{b}$
{\em equally}.
The newly populated level does not have a dominant effect, and no
significant charge imbalance is involved.
Therefore, no energy level locking occurs in quasi-two-dimensional
systems.

In quantum mechanical systems, the effect of  ``energy level
anticrossing'' is very common.
Anticrossings occur because quantum hybridization between the
sub-systems becomes important in near-degenerate situations.
The hybridization opens an energy gap, lifting the incipient level
degeneracy.
Energy level locking is the opposite of anticrossing; i.e., instead of
nearly degenerate energy levels ``repelling'' each other, they lock
together.
The characteristic differences between ancrossings and lockings are
outlined in
Table~\ref{table1}.

In a particular system, level anticrossings and lockings may coexist.
The resultant electronic structure depends on the competition
between these two effects.
In most situations, the effect of energy level locking is very weak.
In the system of two parallel quantum wires, we have demonstrated
by numerical calculations that the energy level locking can be the
dominant effect.

\section{Experimental Implications}
\label{sec:4}
The theoretical results presented above are based on a specific model
of quantum wires.
The experimental realizations of quantum wires are more
complicated systems, with the lateral confinement usually achieved
by means of gates \cite{Thornton,Zhang} rather than the ribbons of
positive charge that we have considered here.
However, the physical mechanism of level locking relies on the
electronic density of states singularity that is common to all
Fermionic parallel quantum wires, irrespective of the method of
confinement.
Another complication is that at present only short quantum wires
(known as ballistic constrictions) are of sufficiently high quality for
experimental studies of energy level locking to be feasible.
In gated parallel constrictions \cite{Smith,Simpson}, the electrons
confined between the gates share the same Fermi energy with the
electron reservoirs of source and drain.
A charge imbalance can be easily achieved by transferring electrons
from or to the reservoirs.
Moreover, the lateral confinement of electrons in the gated
constrictions tends to be stronger than in the Coulomb-confined
systems, so that the quantum hybridization that competes with level
locking should be less important.
Therefore, energy level locking may also occur in gated systems of
similar parallel constrictions.

In a pioneering measurement for a gated system of two parallel
constrictions, Smith {\it et al.}\ \cite{Smith} found that the total
conductance shows successive double steps of $4e^{2}/h$.
The authors suggested that these double steps of conductance result
from non-random alignments of the wire subbands.
The $4e^{2}/h$ double steps have also been observed in
measurements of other similar structures \cite{Hwang,Schmidt}.
In a recent experimental study, however, Simpson {\it et al.}\
\cite{Simpson} compared the total conductance of two quantum
constrictions to the sum of the two individual conductances, but
found no evidence of simultaneous subband depopulation.
Thus the experimental situation at present is unclear.

It is well-known \cite{Ulloa,van Wees,Wharam} that, in an ideal one-
dimensional system, the conductance is quantized, given by
\begin{equation}
G = {2 e^{2} \over h} N_{p},
\label{conductance}
\end{equation}
where $N_{p}$ is the number of populated subbands.
If energy level locking occurs in a gated system that contains two
similar parallel constrictions, while tuning the gate voltage, the Fermi
energy should cross a pair of transverse levels almost at the same
time.
(Notice that {\em exact} energy degeneracies do not occur in
one-dimensional-systems.\ \cite{Landau}) That is, $N_{p}$ changes
by 2 each
time, and, therefore, $G$ should show the double steps of $4e^{2}/h$.

One should note, however, that the curve of conductance vs gate
voltage always has considerable sloping regions between adjacent
plateaux.\ \cite{van Wees,Wharam} If the sloping regions of the
individual
conductances corresponding to the two constrictions overlap
partially, the
total conductance of the system presents a double step as well. It
turns out
that there is a quite high probability that the total conductance
shows a
double step, even if the corresponding levels associating with the two
constrictions are well separated. Therefore, the occurrence of double
steps in
the total conductance curve does not necessarily mean that two
transverse
levels line up exactly.

Considering that one can now tune the gate voltages independently
\cite{Simpson,Feng,Taylor,Yang}, we suggest another way to search
experimentally for energy level locking.
When one tunes one side gate voltage while fixing the other, the
widths of plateaux of the total conductance also change.
Based on our theoretical studies, we know that the locked levels tend
to stay together.
This feature should make the width of a plateau insensitive to the
tuning gate voltage {\em when the plateau width is close to the
maximum}.
This is because the maximum width of the plateau corresponds to the
smallest separation of the two levels, which is the situation of energy
level locking.
Experiments tuning one side gate voltage have been carried out by
Simpson {\em
et al}. \cite{Simpson} However, the voltage steps taken were too
large for this
test for energy level locking to be applied to the published data.
Further
experimental measurements would therefore be of interest.

Besides the transport properties, other measurements, such as the
excitation spectrum, can also in principle be used to verify the
existence of energy level locking in the parallel ballistic constrictions.

In conclusion, we have presented a theoretical study demonstrating
that energy level locking should occur between similar parallel
quantum wires.
It is driven by a charge imbalance associated with the onset of filling
of transverse energy levels with electrons.
This novel phenomenon is qualitatively different from the
anticrossing behavior that is typical of nearly degenerate energy
levels in quantum systems.
Our results should stimulate further experimental and theoretical
studies of this phenomenon.

\acknowledgements
We wish to thank D. Loss, M. Thewalt, and H. Trottier for stimulating
discussions. This work was supported by the National Sciences and
Engineering Research Council of Canada and the Center for Systems
Science at Simon Fraser University.

\figure{Schematic drawing of two parallel Coulomb-confined
quantum wires.
\label{model}}

\figure{The calculated transverse energy levels of two similar
parallel quantum
wires (solid lines) and Fermi energy (dotted line) vs the uniform
donor density
$\sigma$.  Energy levels are labelled A and B according to which
wire they
belong principally to. Arrows indicate anticrossings. Model
parameters are
$w_{a}$ = 190~nm, $w_{b}$ = 200~nm, $s$ = 200~nm,
 and $d$ = 20~nm.
\label{similar}}

\figure{Schematic energy level structure of a pair of parallel wires.
The horizontal axis is the longitudinal wave vector $k$ and the
vertical axis is the subband energy $E_{lk}$.
The parabolic curves are the four lowest subbands.
Level~3 is filled while level~4 is empty.
\label{parabolic}}

\figure{Calculated electron number ratio $r$ between the wires.
$r_{0} = w_{a}/w_{b}$ = 0.95 corresponds to perfect charge balance.
The solid line is a guide to the eye.
\label{transfer}}

\figure{Energy levels (solid lines) and Fermi energy (dotted line) vs
the
inter-wire separation $s$. Parameters are $w_{a}$ = 190~nm,
$w_{b}$ = 200~nm,
$d$ = 20~nm, and $\sigma_{a} = \sigma_{b} = {\rm 2.0 \times
10^{10}~cm^{2} }$.
The right side solid lines and the dashed lines correspond to the
energy levels
of isolated wire A and B, respectively.
\label{wireseparation}}

\figure{Wavefunctions of the four lowest levels when $s$ = 100~nm.
Other
parameters are the same as those in Fig.\ \ref{wireseparation}.
\label{wavefunction}}

\figure{The electronic structures of dissimilar parallel quantum
wires.
$w_{a}$ = 100~nm and $w_{b}$ = 200~nm, and other parameters are
$s$ = 200~nm and $d$ = 20~nm.
The levels are labelled A or B according to which wire they primarily
belong
to.
\label{dissimilar1}}

\figure{$\sigma_{b}$ is varied while $\sigma_{a}$ is fixed at ${\rm
1.5 \times
10^{10}~cm^{2} }$. The parameters are $w_{a}$ = 190~nm, $w_{b}$ =
200~nm, $s$ =
200~nm, and $d$ = 20~nm. Levels are labelled A or B according to
which wire
they primarily belong to.
\label{dissimilar2}}

\begin{table}
\caption{A comparison of the characteristic differences between
anticrossings and
lockings.}
\begin{tabular}{cc}
Anticrossings & Lockings \\ \hline
single-particle effect & many-particle effect \\
opening a gap & reducing the gap \\
caused by wave function overlap & caused by a charge imbalance
	\\
level sequence switches  &no level sequence switches	\\
occurs in all dimension & occurs in one dimension
\end{tabular}
\label{table1}
\end{table}


\begin{references}
\bibitem{Ulloa} For recent reviews see S. E. Ulloa, A. MacKinnon, E.
Casta\~{n}o, and G. Kirczenow, {\it From Ballistic Transport to
Localization}, edited by P. T. Landsberg, Handbook of Semiconductors
Vol. I (North-Holland, Amsterdam, 1992); C. W. J. Beenakker and H.
van Houten, {\it Quantum Transport in Semiconductor
Nanostructures}, edited by H. Ehrenreich and D. Turnbull, Solid State
Physics, Advances in Research and Applications Vol. 44 (Academic
Press, San Diego, 1991).
\bibitem{Roukes} For a review, see M. L. Roukes, T. J. Thornton, A.
Scherer, J. A. Simmons, B. P. van der Gaag, and E. D. Beebe, in {\it
Science and Engineering of 1- and 0-Dimensional Semiconductors},
edited by S. P. Beaumont and C. M. Sotomayer-Torres (Plenum,
London, 1990).
\bibitem{Lai} W. Y. Lai and S. Das Sarma, Phys. Rev. {\bf B33}, 8874
(1986).
\bibitem{Laux-86} S. E. Laux and F. Stern, Appl. Phy. Lett. {\bf 49},
91 (1986).
\bibitem{Laux-88} S. E. Laux, D. J. Franck, and F. Stern, Surf. Sci. {\bf
196}, 101 (1988).
\bibitem{Davies} J. H. Davies, Semicond. Sci. Technol. {\bf 3}, 995
(1988).
\bibitem{Glazman} L. I. Glazman and I. A. Larkin, Superlatt.
Microstruct. {\bf 6}, 32 (1988).
\bibitem{Nakamura} A. Nakamura and A. Okiji, J. Phys. Soc. Jpn. {\bf
60}, 1873 (1991).
\bibitem{Sun-93} Y. Sun and G. Kirczenow, Phys. Rev. {\bf B47}, 4413
(1993).
\bibitem{Avishai} Y. Avishai, M. Kaveh, S. Shatz, and Y. B. Band, J.
Phys.: Condens. Matter {\bf 1}, 6907 (1989).
\bibitem{Castano} E. Casta\~{n}o and G. Kirczenow, Phys. Rev. {\bf
B41}, 5055 (1990).
\bibitem{Vasilopoulos} Y. M. Sirenko, P. Vasilopoulos, and I. I. Boiko,
Phys. Rev. {\bf B44}, 10724 (1991); H. C. Tso and P. Vasilopoulos,
Phys. Rev. {\bf B45}, 1333 (1992); I. I. Boiko, P. Vasilopoulos, and V.
I. Sheka, Phys. Rev. {\bf B45}, 135724 (1992).
\bibitem{Ravaioli} U. Ravaioli, T. Kerkhoven, M. Raschke, and A. T.
Galick, Superlatt. Microstruc. {\bf 11}, 343 (1992).
\bibitem{Sun-94} A preliminary account of part of this work has
already been published. See Y. Sun and G. Kirczenow, Phys. Rev. Lett.
{\bf 72}, 2450 (1994).
\bibitem{Hohenberg} P. Hohenberg and W. Kohn, Phys. Rev. {\bf
136}, B864 (1964).
\bibitem{Kohn} W. Kohn and L. J. Sham, Phys. Rev. {\bf 140}, A1133
(1965).
\bibitem{Stern} F. Stern, Phys. Rev. Lett. {\bf 30}, 278 (1973).
\bibitem{Jonson} M. Jonson, J. Phys. {\bf C9}, 3055 (1976).
\bibitem{Tanatar} B. Tanatar and D. M. Ceperly, Phys. Rev. {\bf B39},
5005 (1989).
\bibitem{Thornton} T. J. Thornton, M. Pepper, H. Ahmed, D. Andrews,
and G. J. Davies, Phys. Rev. Lett. {\bf 56}, 1198 (1986).
\bibitem{Zhang} H. Z. Zhang, H. P. Wei, D. C. Tsui, and G. Weimann,
Phys. Rev. {\bf B34}, 5635 (1986).
\bibitem{Smith} C. G. Smith, M. Pepper, R. Newbury, H. Ahmed, D. G.
Hasko, D. C. Peacock, J. E. F. Frost, D. A. Ritchie, G. A. C. Jones, and G.
Hill, J. Phys.: Condens. Matter {\bf 1}, 6763 (1989).
\bibitem{Simpson} P. J. Simpson, D. R. Mace, C. J. B. Ford, I. Zailer, M.
Pepper, D. A. Ritchie, J. E. F. Frost, M. P. Grimshaw, and G. A. C. Jones,
Appl. Phy. Lett. {\bf 63}, 3191 (1993).
\bibitem{Hwang} S. W. Hwang, J. A. Simmons, D. C. Tsui, and M.
Shayegan, Phys. Rev. {\bf 44}, 13497 (1991).
\bibitem{Schmidt} P. E. Schmidt, M. Okada, K. Kosemura, and N.
Yokoyama, Jpn. J. Appl. Phys. {\bf 30}, L1921 (1991).
\bibitem{van Wees} B. J. van Wees, H. van Houten, C. W. J. Beenakker,
J. G. Williamson, L. P. Kouwenhoven, D. van der Marel, and C. T. Foxon,
Phys. Rev. Lett. {\bf 60}, 848 (1988).
\bibitem{Wharam} D. A. Wharam, T. J. Thorton, R. Newbury, M.
Pepper, H. Ahmed, J. E. F. Frost, D. G. Hasko, D. C. Peacock, D. A.
Ritchie, and G. A. C. Jones, J. Phys. C: Solid State Phys. {\bf 21}, L209
(1988).
\bibitem{Landau} In one-dimensional quantum systems, exact
energy level degeneracies do not occur. See L. D. Landau, E. M.
Lifshitz and L. P. Pitaevskii, {\it Quantum Mechanics, Non-Relativistic
Theory} (Pergamon, New York, 1986), 3rd ed., Sec. 21.
\bibitem{Feng} Y. Feng, A. S. Sachrajda, R. P. Taylor, J. A. Adams, M.
Davies, P. Zawadzki, P. T. Coleridge, D. Landheer, P. A. Marshall, and R.
Barber, Appl. Phy. Lett. {\bf 63}, 1666 (1993).
\bibitem{Taylor} R. P. Taylor, J. A. Adams, M. Davies, P. A. Marshall,
and R. Barber, J. Vac. Sci. Technol.  {\bf B11}, 628 (1993).
\bibitem{Yang} S. Yang, M. J. Berry, A. S. Adourian, R. M. Westervelt,
and A. C. Gossard, Bull. Am. Phys. Soc. {\bf 37}, 70
(1994).
\end{references}
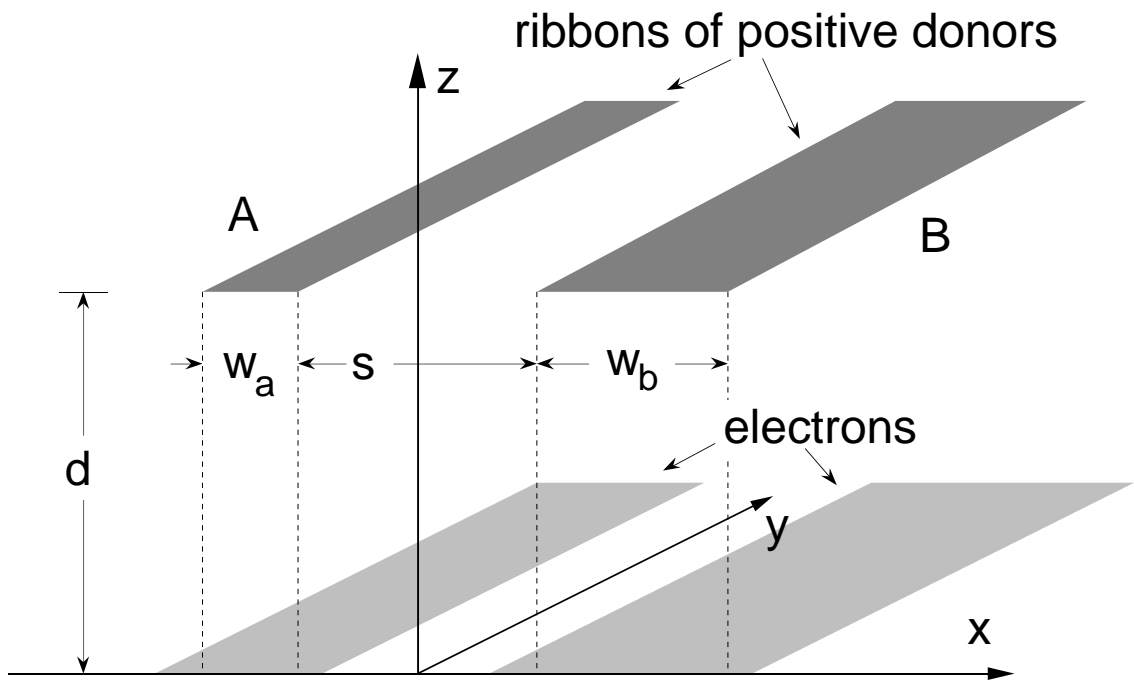
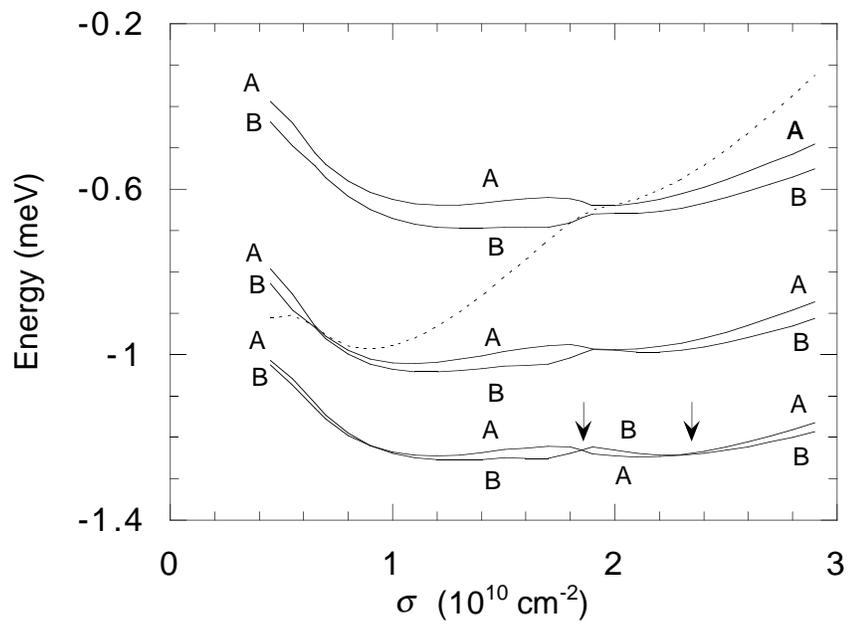
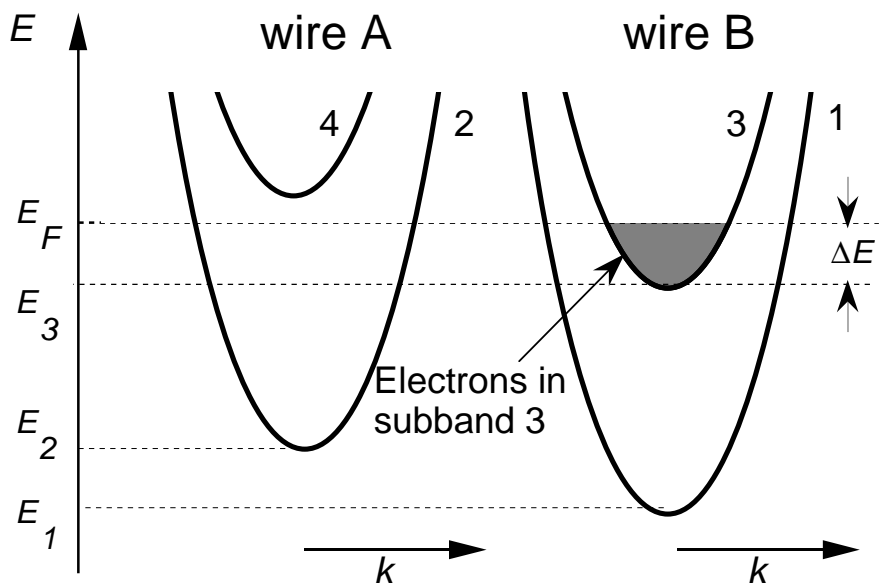
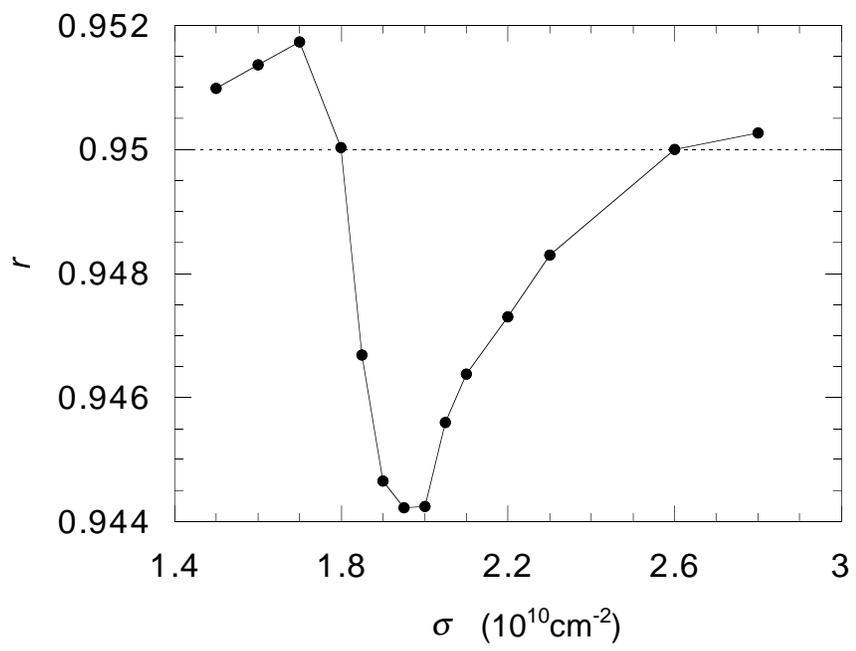
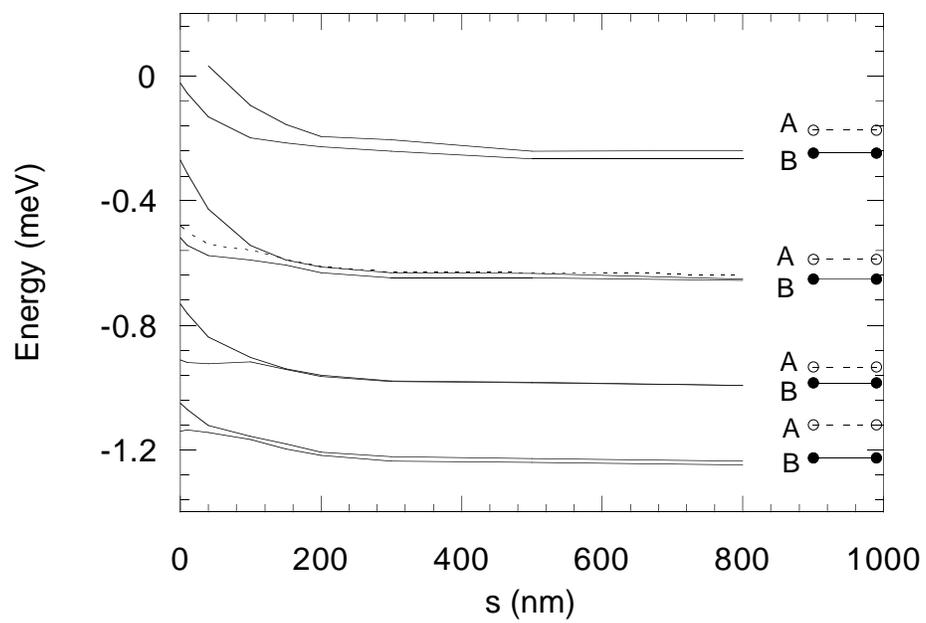
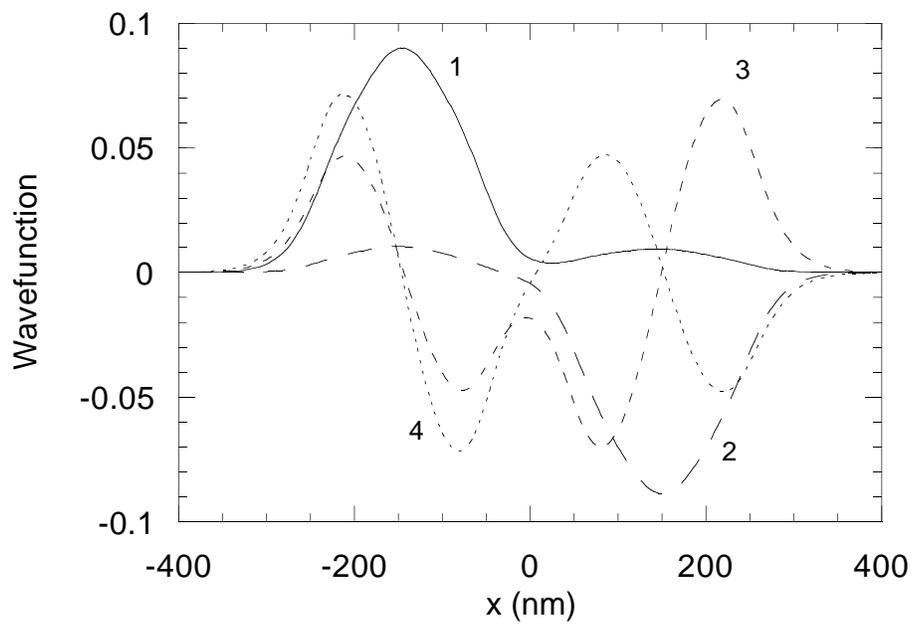
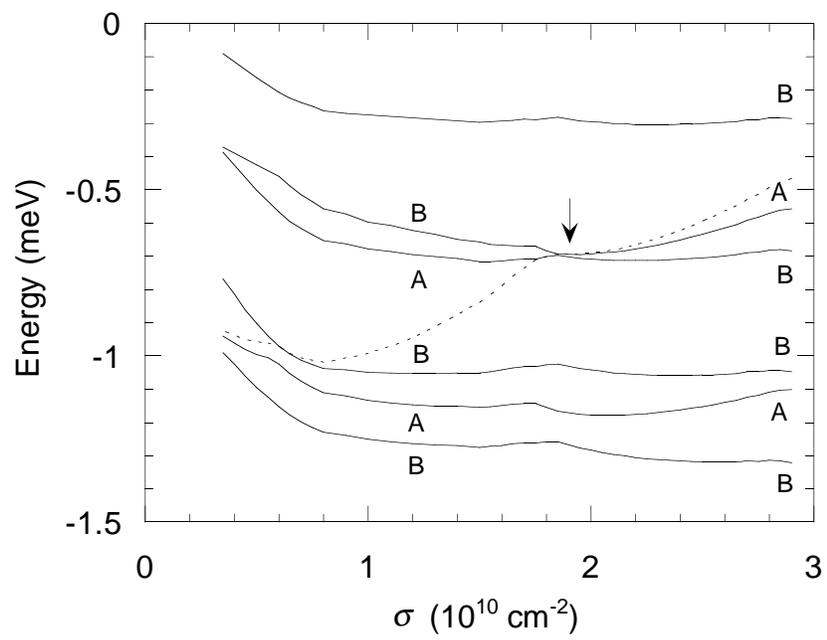
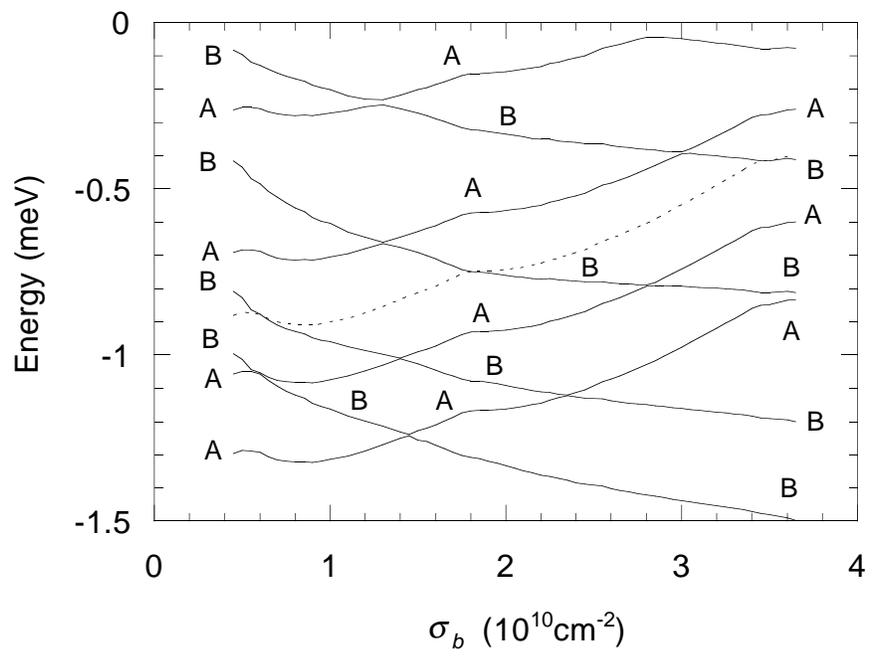
\end{document}